\documentclass[aps, prl, reprint, longbibliography]{revtex4-2}

\usepackage[utf8]{inputenc}
\usepackage[english]{babel}
\usepackage[colorlinks, citecolor=blue]{hyperref}
\usepackage{graphicx}
\usepackage{amssymb, amsmath}
\usepackage{bm}
\usepackage{siunitx}
\usepackage{mathptmx}
\usepackage{mathtools}
\usepackage{verbatim}
\usepackage{anyfontsize}
\usepackage{multirow}

\begin{document}

\title{Tuneable Gaussian entanglement in levitated nanoparticle arrays}

\author{Anil Kumar Chauhan}
\email{anil.kumar@upol.cz}
\affiliation{Department of Optics, Palack\'y University, 17. listopadu 1192/12, 77146 Olomouc, Czech Republic}

\author{Ond\v{r}ej \v{C}ernot\'{i}k}
\email{ondrej.cernotik@upol.cz}
\affiliation{Department of Optics, Palack\'y University, 17. listopadu 1192/12, 77146 Olomouc, Czech Republic}

\author{Radim Filip}
\email{filip@optics.upol.cz}
\affiliation{Department of Optics, Palack\'y University, 17. listopadu 1192/12, 77146 Olomouc, Czech Republic}

\date{\today}

\begin{abstract}
	Nanoparticles trapped in optical tweezers emerged as an interesting platform for investigating fundamental effects in quantum physics. The ability to shape the optical trapping potential using spatial light modulation and quantum control of their motion using coherent scattering to an optical cavity mode predispose them for emulating a range of physical systems and studying quantum phenomena with massive objects. To extend these capabilities of levitated nanoparticles to quantum many-body systems, it is crucial to develop feasible strategies to couple and entangle multiple particles either directly or via a common optical bus. Here, we propose a variable and deterministic scheme to generate Gaussian entanglement in the motional steady state of multiple levitated nanoparticles using coherent scattering to multiple cavity modes. Coupling multiple nanoparticles to a common optical cavity mode allows cooling of a collective Bogoliubov mode to its quantum ground state; cooling multiple Bogoliubov modes (enabled by trapping each particle in multiple tweezers such that each tweezer scatters photons into a separate cavity mode) removes most thermal noise, leading to strong entanglement between nanoparticles. We present numerical simulations for three nanoparticles showing great tuneability of the generated entanglement with realistic experimental parameters. Our proposal thus paves the way towards creating complex quantum states of multiple levitated nanoparticles for advanced quantum sensing protocols and many-body quantum simulations.
\end{abstract}

\maketitle


Optomechanics with levitated nanoparticles has emerged as an attractive platform for sensing~\cite{Gieseler2013,Hebestreit2018,Ahn2020}, thermodynamics~\cite{Dechant2015,Gieseler2018,Debiossac2020}, and tests of fundamental physics~\cite{Arndt2014,Moore2021}. Optical, electrical, and magnetic trapping allows a broad range of nanoparticles to be stably trapped in both harmonic and anharmonic potentials, and the isolation from solid-state substrates and surfaces results in exceptional isolation from environment, enabling ultra-high quality factors of their centre-of-mass, librational, and rotational modes~\cite{Millen2020,gonzalez2021,Stickler2021}. Efficient control techniques can be used to prepare nonclassical states of nanoparticle motion, opening the way to practical applications of levitated systems in quantum technologies. With recent demonstrations of the quantum regime of nanoparticle motion~\cite{Tebbenjohanns2020,Magrini2021,Tebbenjohanns2021,Magrini2022X,Militaru2022X}, it is only a matter of time before we see the realisation of more complex quantum states and further groundbreaking experiments.

Particularly optical levitation and control offers a great promise as it capitalises on existing theoretical and experimental techniques of cavity optomechanics~\cite{Aspelmeyer2014,Barzanjeh2022} and atomic physics~\cite{Chang2009,Romero-Isart2011a}. The large dipole induced by the trapping field in a dielectric nanoparticle behaves just like the intrinsic dipole of an individual atom; the much larger size and mass of the nanoparticles opens a new regime of quantum sensing experiments and enables tests of the limits of quantum mechanics~\cite{Weiss2021}. A remarkable example of levitated optomechanics borrowing from atomic physics is coherent scattering, in which optomechanical interaction is mediated by the nanoparticle scattering tweezer photons into an empty cavity mode~\cite{Gonzalez-Ballestero2019}. Initially developed as a method for cooling the motion atoms, ions, and molecules~\cite{Vuletic2000,Leibrandt2009}, it has now been used to the same effect in levitated optomechanics~\cite{delic2019,Windey2019,delic2020}, for demonstrating optomechanical strong coupling~\cite{RiosSommer2021}, and proposals exist to use the same mechanism for creating mechanical squeezing~\cite{Cernotik2020,Kustura2022} and generating entanglement~\cite{anil2022,Rakhubovsky2020,Rudolph2020,Brandao2021,Li2021X}.

With the tremendous progress in the past years in controlling the motion of levitated nanoparticles, it is natural to expect that experiments will soon turn to using multiple nanoparticles trapped in regular geometric structures and forming optically levitated nanoparticle crystals. Following the progress with trapped atoms and atomic lattices, we can expect arrays of levitated nanoparticles being used for advanced quantum sensing protocols~\cite{Sewell2014,Zhou2020}, quantum simulations~\cite{Bernien2017,Arguello-Luengo2019,Wintersperger2020,Bluvstein2021} and information processing~\cite{Ebadi2021,Bluvstein2022}, building novel metamaterials~\cite{Shahmoon2017,Bekenstein2020,Rui2020}, and other applications~\cite{Peyronel2012,McConnell2015}. In a bottom-up approach where more and more complex structures are built starting from simpler, few-particle systems, the most pressing theoretical questions are the following: How can we efficiently control the motional states in these multiparticle systems? And what nonclassical effects can we expect to observe? These answers have been studied intensively for two-particle systems~\cite{anil2022,Penny2021X,DeBernardis2022,rieser2022} but scaling to larger nanoparticle arrays remains an open question.

Here, we take the first step in this direction by proposing a scheme to generate deterministic Gaussian entanglement in the steady state of multiple levitated nanoparticles. Coherent scattering of tweezer photons from multiple nanoparticles into a common cavity mode allows us to cool their collective Bogoliubov mode to its quantum ground state~\cite{anil2022}. Employing multiple tweezers per particle to scatter photons into separate cavity modes enables cooling of multiple Bogoliubov modes, efficiently removing thermal fluctuations from the multiparticle state and creating strong multipartite entanglement between them. Crucially, the strategy generates entanglement deterministically in the steady state. The resulting state is Gaussian, limiting its applications in quantum experiments and technology applications. However, complementing our scheme with nonlinear potentials for trapping the particles~\cite{Siler2018,Gieseler2013} or photon counting~\cite{Rudolph2020,Riedinger2016} would allow us to move beyond the Gaussian regime and unlock the full potential of optically levitated nanoparticles in quantum physics.

We focus on the test case of three levitated nanoparticles in numerical simulations, which allows us to fully characterise their resulting entanglement via the positive partial transpose (PPT) criterion as it is a necessary and sufficient condition for separability of any bipartition of two and three modes in the Gaussian regime~\cite{Werner2001}. We demonstrate great versatility of the scheme that allows us to change the number of entangled bipartitions by changing the structure of the Bogoliubov modes and their coupling to the cavity modes. While we focus primarily on bipartite entanglement as it can be efficiently characterized and quantified using the PPT criterion, we also demonstrate that it is possible to create genuine tripartite entanglement in state-of-the-art levitated systems. Our work thus further confirms the potential of coherent scattering for controlling the motion of levitated nanoparticles and presents a viable approach to investigating quantum many-body dynamics in nanoparticle arrays.

\section{Results}

\subsection{Model and dynamics}

To set the stage, we begin with one particle in one tweezer coherently scattering photons into an empty cavity mode. Depending on the detuning between the tweezer and cavity mode, the scattering can give rise to beam-splitter coupling (for tweezer frequency $\omega_{\rm tw}$ smaller than cavity frequency $\omega_c$ by the mechanical frequency $\omega_m$, $\omega_{\rm tw} = \omega_c-\omega_m$), $H = g_-(a^\dagger b+b^\dagger a)$, or two-mode squeezing interaction (for $\omega_{\rm tw} = \omega_c+\omega_m$), $H = g_+(ab+a^\dagger b^\dagger)$~\cite{gonzalez2021}. Here $a$ and $b$ are the annihilation operators of the cavity field and the mechanical mode and $g_\pm$ are the interaction strengths of the two-mode squeezing and beam-splitter interaction. The beam-splitter interaction removes energy from the mechanical mode, leading to its cooling~\cite{delic2020}, while two-mode squeezing generates photon--phonon pairs, generating entanglement between the cavity and mechanical modes.

\begin{figure}
	\centering
	\includegraphics[width=\linewidth]{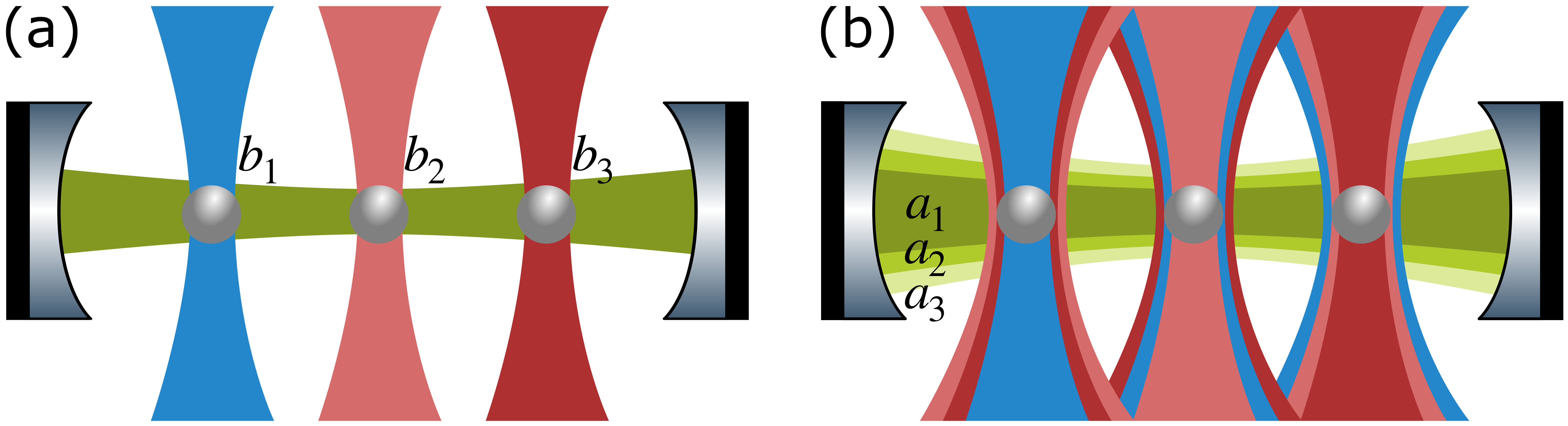}
	\caption{\label{fig_1}
	Illustration of cooling collective Bogoluibov modes by applying multiple tweezers per particle. (a) Cooling of one Bogoliubov mode of three particles created by a combination of two-mode squeezing (blue tweezer) and beam-splitter (red tweezer) interactions. The relative strength of individual particle modes in the Bogoliubov modes is set by the relative intensity of the tweezers (light or dark red for particles $b_2$ and $b_3$). (b) Cooling of three Bogoliubov modes by coupling each to a different cavity modes (dark to light green).
	}
\end{figure}

If multiple particles are coupled to the same cavity mode by a combination of beam-splitter and two-mode squeezing interactions (see Fig.~\ref{fig_1}(a)), their collective Bogoliubov mode is coupled to the cavity and can be efficiently cooled by the optomechanical interaction,
\begin{align}\label{eq:Bog1}
\begin{split}
	H &= \sum_{j=1}^N g_{-,j}(a^\dagger b_j + b_j^\dagger a) + g_{+,j}(ab_j+a^\dagger b_j^\dagger)\\
	&= G(a^\dagger\beta+\beta^\dagger a),
\end{split}
\end{align}
where we assume that each mechanical mode $b_j$ is coupled to the cavity by either beam-splitter or two-mode squeezing interaction (but not both at the same time) and $g_{\pm,j}\in\mathbb{R}$ without loss of generality.
The Bogoliubov mode~\cite{Woolley2014} is given by
\begin{equation}
	\beta = \frac{1}{G}\left(\sum_j g_{-,j}b_j+g_{+,j}b_j^\dagger\right)
\end{equation}
and we have introduced the effective optomechanical coupling
\begin{equation}
	G^2 = \sum_j g_{-,j}^2-g_{+,j}^2;
\end{equation}
for dynamical stability, we require that the sum of all beam-splitter interactions be stronger than the two-mode squeezing interactions, which guarantees $G^2>0$, ensuring that the Hamiltonian~\eqref{eq:Bog1} is Hermitian.

The beam-splitter interaction between the Bogoliubov mode and the cavity field cools down the collective mechanical mode by coupling it to the zero-temperature optical bath. The final occupation of the Bogoliubov mode (assuming, for simplicity, equal thermal occupation and damping rates for all particles $b_j$) then scales with the total cooperativity, $n_{\rm eff} = \langle\beta^\dagger\beta\rangle \sim 1/C$, $C = 4G^2/\kappa\gamma\bar{n}$, where $\kappa$ and $\gamma$ are the optical and mechanical damping rates and $\bar{n}$ is the thermal occupation of the mechanical modes. In the regime of strong cooperativity, $C>1$, the effective occupation reaches $n_{\rm eff}<1$ and the Bogoliubov mode is effectively in the quantum ground state~\cite{Marquardt2007,Wilson-Rae2007}.

With the Bogoliubov mode cooled to its quantum ground state, the variances of its quadrature operators $x_\beta = (\beta + \beta^\dagger)/\sqrt{2}$, $p_\beta = -i(\beta-\beta^\dagger)/\sqrt{2}$ are close to the vacuum level (up to the small residual thermal noise $n_{\rm eff}<1$). In terms of the particle modes $b_j$, these quadratures can be expressed as
\begin{equation}\label{eq:collective}
	x_\beta = \frac{1}{G}\sum_j (g_{-,j}+g_{+,j})x_j,\quad
	p_\beta = \frac{1}{G}\sum_j (g_{-,j}-g_{+,j})p_j,
\end{equation}
where the quadrature operators $x_j,p_j$ are defined in full analogy with the quadratures of the Bogoliubov mode. The modes thus remain separable since generation of entanglement would require the collective quadratures~\eqref{eq:collective} to be squeezed below the vacuum level~\cite{Duan2000}. However, the classical correlations, generated between the particles by the passive interaction~\eqref{eq:Bog1}, can be directly observed by measuring the collective quadratures~\eqref{eq:collective} and demonstrating that their variance is smaller than the thermal noise level of the particle bath. Such an experiment would present a significant milestone towards quantum mechanics of nanoparticle arrays as it would clearly demonstrate efficient manipulation of collective motional modes of multiple nanoparticles.

Cooling one Bogoliubov mode via the Hamiltonian~\eqref{eq:Bog1} is not sufficient to create entanglement between the particles since the remaining $N-1$ normal modes (which, together with the Bogoliubov mode $\beta$, form an orthonormal basis for the nanoparticle array) of the mechanical subsystem are in a thermal state with large thermal occupation $\bar{n}$. To cool these modes to the ground state, we can use multiple tweezers per particle which scatter photons into separate cavity modes with suitable tweezer detunings (Fig.~\ref{fig_1}(b)),
\begin{align}\label{m_h3}
\begin{split}
	H &= \sum_{k=1}^N G_k(a_k^\dagger\beta_k + \beta_k^\dagger a_k)\\
	&= \sum_{j,k=1}^N g_{-,jk}(a_k^\dagger b_j+b_j^\dagger a_k) + g_{+,jk}(a_kb_j+a_k^\dagger b_j^\dagger).
\end{split}
\end{align}
The full dynamics of this multimode manybody system can be described by the linear Langevin equations
\begin{subequations}\label{eq:Langevin}
\begin{align}
	\dot{a}_k &= -i\sum_j(g_{-,jk}b_j+g_{+,jk}b_j^\dagger) -\frac{\kappa_k}{2}a_k + \sqrt{\kappa_k}a_{k,{\rm in}},\\
	\dot{b}_j &= -i\sum_k(g_{-,jk}a_k+g_{+,jk}a_k^\dagger) -\frac{\gamma_j}{2}b_j + \sqrt{\gamma_j}b_{j,{\rm in}},
\end{align}
\end{subequations}
where $\kappa_k$ and $\gamma_j$ are the decay rates of the cavity modes $a_k$ and mechanical modes $b_j$, respectively.
The optical input noise operators $a_{k,{\rm in}}$ fulfil the usual correlations $\langle a_{k,{\rm in}}(t) a^\dag_{k,{\rm in}}(t')\rangle = \delta(t-t')$ and the thermal noise operators of the mechanical modes follow $\langle b_{j,{\rm in}}(t) b^\dag_{j,{\rm in}}(t')\rangle = (2 \bar{n} + 1) \delta(t-t')$; we assume for simplicity that all mechanical modes have the same average occupation $\bar{n}$.

\subsection{Entanglement generation and classification}

The dynamics described by Eqs.~\eqref{eq:Langevin} can be solved in terms of the covariance matrix of the quadrature operators $X_k = (a_k+a_k^\dagger)/\sqrt{2}$, $Y_k = -i(a_k-a_k^\dagger)/\sqrt{2}$, $x_j = (b_j+b_j^\dagger)/\sqrt{2}$, $p_j = -i(b_j-b_j^\dagger)/\sqrt{2}$ (see Methods). The entanglement properties of the Gaussian state of the nanoparticles can be studied using the PPT criterion applied to the part of the covariance matrix describing the mechanical modes. Since the PPT criterion is necessary and sufficient only for $1\times M$ Gaussian states, we focus on $N=3$ particles in the rest of this article. This allows us to study entanglement in every pair of particles and in every bipartition of the whole mechanical system using the PPT criterion; for a four-particle system, we would also need to analyse $2\times2$ bipartitions, for which the PPT criterion is only sufficient but not necessary~\cite{Werner2001}.

For three particles, we need three cavity modes for efficient cooling,
\begin{equation}\label{M_b1}
	H = \sum_{k=1}^3 G_k(a_k^\dagger \beta_k+\beta_k^\dagger a_k).
\end{equation}
In the following, we consider mechanical Bogoliubov modes $\beta_k$ parametrized as follows (see also Fig.~\ref{fig_1}(b)):
\begin{subequations}\label{M_b2}
\begin{align}
	\beta_1 &= \lambda_1 b_1^\dag +\lambda_2 b_2 +\lambda_3 b_3, \\
	\beta_2 &= \lambda_3 b_1 +\lambda_1 b_2^\dag +\lambda_2 b_3, \\
	\beta_3 &= \lambda_2 b_1 +\lambda_3 b_2 +\lambda_1 b_3^\dag,
\end{align}
\end{subequations}
where normalization of these modes, $[\beta_j,\beta_j^\dagger]=1$, dictates $-\lambda_1^2+\lambda_2^2+\lambda_3^2=1$. These modes are not fully orthogonal (we have $[\beta_1,\beta_2] = \lambda_1(\lambda_2-\lambda_3)$, $[\beta_1,\beta_2^\dagger] = \lambda_2\lambda_3$, and similar expressions hold for the other combinations) but their permutation symmetry allows insight into their entanglement properties that would be difficult with more general, orthogonal Bogoliubov modes. At the same time, the non-orthogonality of the Bogoliubov modes reduces the total cooling rate and the final thermal occupation of the mechanical modes~\cite{Liu2022} (see also Methods). Nevertheless, we use these Bogoliubov modes only as a tool to understand the system dynamics and resulting entanglement; we are ultimately interested in entanglement between the bare particle modes $b_j$.

\begin{figure}
	\centering
	\includegraphics[width=\linewidth]{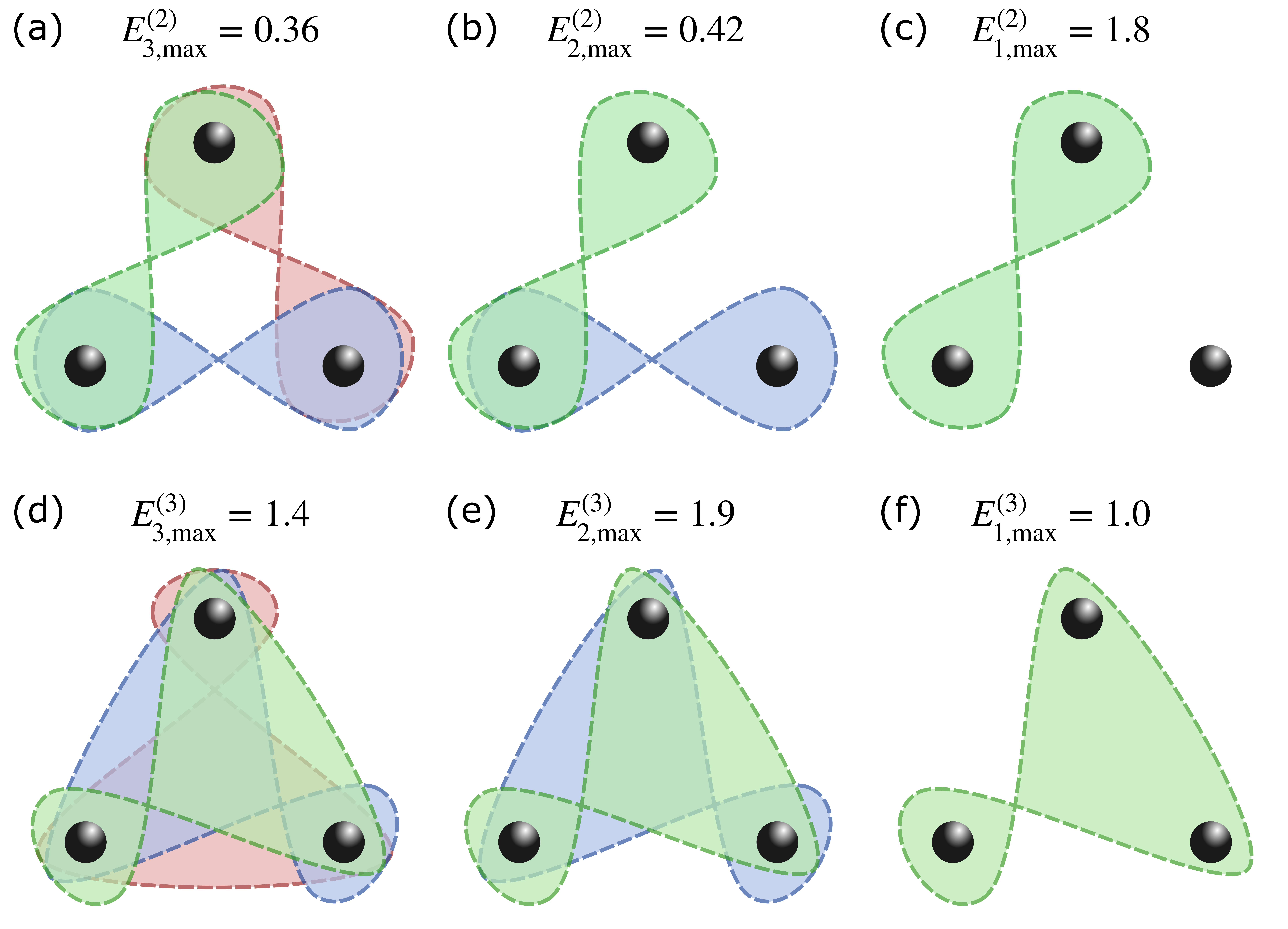}
	\caption{\label{fig:MainRes}
	Entanglement structure in a three-particle system. Dyadic entanglement (between two particles with the third particle traced out) can be observed in (a) all three pairs, (b) two pairs, or (c) one pair. Triadic entanglement (in the full three-particle covariance matrix with one bipartition containing one particle and the other bipartition consisting of the remaining two particles) can be observed in (d) all three bipartitions, corresponding to genuine tripartite entanglement, (e) two bipartitions, or (f) one bipartition. For each configuration, the maximum achievable entanglement (as quantified by logarithmic negativity) is shown (for assumed experimental parameters, see the caption of Fig.~\ref{fig2}). The case of all three bipartitions separable is not shown.}
\end{figure}

We study both pairwise entanglement (i.e., entanglement in a pair of nanoparticles obtained after tracing out the third nanoparticle where each bipartition is formed by one particle), which we call \emph{dyadic} from now on, and entanglement in the whole three-particle system (in which one bipartition is formed by one particle and the other bipartition by the remaining two particles), which we call \emph{triadic}.
For both dyadic and triadic entanglement, the following situations can arise: (i) all three bipartitions are entangled, (ii) two bipartitions are entangled, (iii) only one bipartition is entangled, (iv) all bipartitions are separable (see also Fig.~\ref{fig:MainRes}).
By controlling the structure of the Bogoliubov modes (the parameters $\lambda_j$) and their coupling to the cavity modes (the coupling rates $G_k$), all these scenarios can be realised.

We quantify the generated Gaussian entanglement using logarithmic negativity~\cite{Vidal2002}.
To capture the structure of the generated entanglement as described in the previous paragraph and shown schematically in Fig.~\ref{fig:MainRes}, we proceed as follows:
We calculate the logarithmic negativity for all three bipartitions from the mechanical covariance matrix and sort them in descending order, $E_1 > E_2 > E_3$.
We then define the figures of merit
\begin{subequations}
\begin{align}
	E_3^{(j)} &= \sqrt[3]{E_1E_2E_3},\\
	E_2^{(j)} &= \sqrt{E_1E_2},\\
	E_1^{(j)} &= E_1,
\end{align}
\end{subequations}
where $j$ determines the arity of the entanglement ($j=2$ for dyadic and $j=3$ for triadic entanglement),
corresponding, respectively, to the cases plotted in Fig.~\ref{fig:MainRes}(a,d), (b,e), and (c,f).
To ensure that only two bipartitions are entangled (only one bipartition is entangled), we further require that $E_2^{(j)}\gg E_3$ ($E_1^{(j)}\gg E_2$); otherwise, we set $E_2^{(j)} = 0$ ($E_1^{(j)} = 0$).
Nonzero value of $E_3^{(j)}$ then clearly signifies that all bipartitions of given arity are entangled (corresponding to Fig.~\ref{fig:MainRes}(a) for $j=2$ and (d) for $j=3$), nonzero value of $E_2^{(j)}$ shows that exactly two bipartitions are entangled (Fig.~\ref{fig:MainRes}(b,e); the condition $E_2^{(j)}\gg E_3$ ensures that there is negligible entanglement in the third bipartition), and nonzero value of $E_1^{(j)}$ demonstrates that only one bipartition is entangled (Fig.~\ref{fig:MainRes}(c,f)).
Defining the relevant figures of merit in terms of a geometric mean of the logarithmic negativities of individual bipartitions ensures that the amounts of entanglement in all bipartitions is comparable, $E_1\sim E_2\sim E_3$ for $E_3^{(j)}$ and $E_1\sim E_2$ for $E_2^{(j)}$.
Since the Bogoliubov modes are permutation symmetric, we can generate entanglement in any bipartition (or any two bipartitions) just by cycling through the coupling rates $G_k$.

\subsection{Dyadic entanglement}

\begin{figure}
	\centering
	\includegraphics[width=\linewidth]{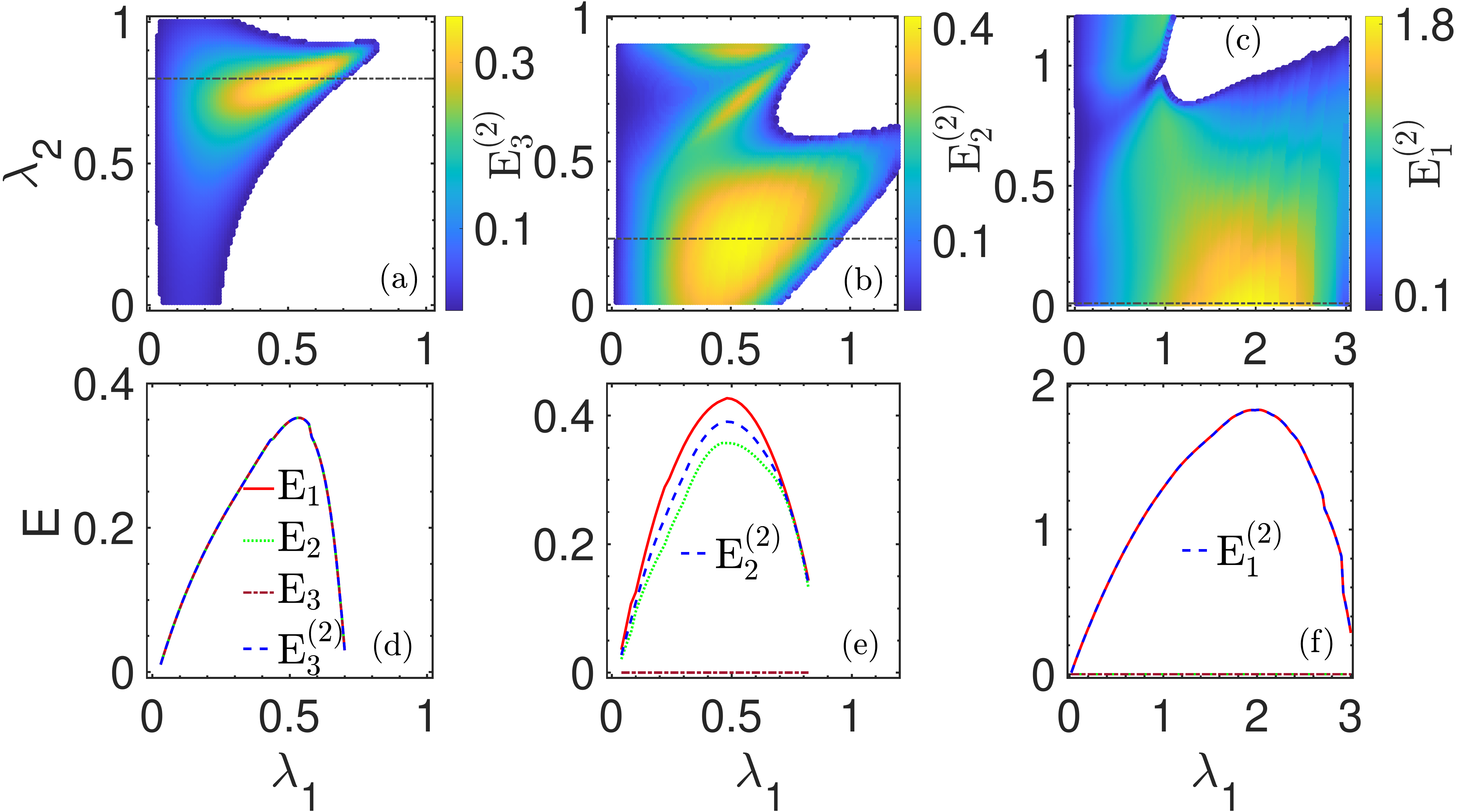}
	\caption{\label{fig2}
Logarithmic negativity quantifying dyadic entanglement. Panels (a)--(c) show numerically optimized entanglement as a function of the Bogoliubov coefficients $\lambda_{1,2}$ (note that $\lambda_3^2=1+\lambda_1^2-\lambda_2^2$ due to normalization). We plot entanglement in (a) all pairs, $E_3^{(2)}$, (b) two pairs, $E_2^{(2)}$, and (c) one pair, $E_1^{(2)}$. Cuts along the black dot-dashed lines through the maximum entanglement are shown in panels (d)--(f) (dashed blue line) together with entanglement in individual pairs (solid red line for $E_1$, dotted green line for $E_2$, and dot-dashed maroon line for $E_3$); note that $E_3=0$ in (e) and (f), and $E_2=0$ in (f). The plots show entanglement in (d) three pairs for $\lambda_2 = 0.8$, (e) two pairs for $\lambda_2 = 0.23$, and (f) one pair for $\lambda_2 = 0.01$.
We assume the mechanical quality factor $Q = \omega_m/\gamma =5\times 10^9$ and mean thermal occupation of $\bar{n}=2\times 10^7$ equal for all mechanical modes and linewidth for all cavity modes $\kappa=0.4$ $\omega_m$. For numerical optimization, the maximum coupling rates are set at $G_k = 0.4\omega_m$.
	}
\end{figure}

We analyse the attainable dyadic entanglement numerically in Fig.~\ref{fig2} against the coefficients $\lambda_{1,2}$ of the Bogoliubov modes; note that owing to normalization, the third coefficient is given by $\lambda_3 = \sqrt{1+\lambda_1^2-\lambda_2^2}$. For each data point, we further optimize the observed entanglement numerically by finding the coupling rates $G_k$ that maximize the given figure of merit. The structure of the Bogoliubov modes, set by the coefficients $\lambda_k$, and their coupling strengths, $G_k$, fully determine the structure of the resulting entanglement, allowing us to create strong entanglement in (a) all pairs, (b) two pairs, and (c) one pair of nanoparticles for experimentally achievable system parameters~\cite{delic2019,Windey2019,delic2020}.

For creating entanglement in all pairs (panel (a)), we assume equal coupling of all Bogoliubov modes, $G_1 = G_2 = G_3$. Due to permutation symmetry of the modes, this choice creates equal entanglement in all three pairs of particles. This is further supported by the line plot in panel (d) where we show the cut through the maximum entanglement (reached for $\lambda_2 = 0.8$, see the black dot-dashed line in (a)) along with the entanglement in all three pairs $E_k$. The smaller maximum (compared to $E_1^{(2)}$ and triadic entanglement) of about $E_{3,{\rm max}}^{(2)} \simeq 0.36$ is caused by tracing out one of the particles when calculating the logarithmic negativity. Each Bogoliubov mode creates strong entanglement in the full three-particle system and so tracing out one of the particles appears as thermal noise, reducing the amount of observable entanglement.

When we choose different coupling rates for the three Bogoliubov modes, $G_1\neq G_2\neq G_3$, the permutation symmetry is broken, allowing us to create entanglement only in selected pairs of particles as shown in panel (b) for entanglement in two pairs. The maximum entanglement, $E_{2,{\rm max}}^{(2)}\simeq 0.42$, is obtained by cooling the three Bogoliubov modes using different rates each, $G_1 > G_2 > G_3$. Due to the large asymmetry in the Bogoliubov coefficients $\lambda_k$ (we have $\lambda_3\gg\lambda_1>\lambda_2$), cooling the first Bogoliubov mode creates primarily strong entanglement between particles $b_1$ and $b_3$. The weaker entanglement between particles $b_1$ and $b_2$ is then enhanced by the cooling of the Bogoliubov mode $\beta_2$. The remaining pair of particles---$b_2$ and $b_3$---can become entangled only through the much weaker coupling of these particles in the mode $\beta_2$ and inefficient cooling of the mode $\beta_3$ which are not strong enough. These points are further illustrated in panel (e) which shows the cut through the maximum $E_{2,{\rm max}}^{(2)}$ (for $\lambda_2=0.23$): The logarithmic negativities in two pairs are large and close to each other (reaching the maxima $E_1\simeq 0.43$, $E_2\simeq 0.41$) while there is no entanglement in the third pair.

Finally, entanglement generation in a single pair of particles is studies in panel (c). In this case, the optimal choice is $\lambda_2\simeq 0$ with strong cooling of only one Bogoliubov mode, $G_1\gg G_2\sim G_3$. This results in strong entanglement in the particles $b_1$ and $b_3$ with no entanglement in the remaining pairs (as shown in the cut for $\lambda_2 = 0.01$ in panel (f)). However, nonzero coupling of the remaining two Bogoliubov modes helps reduce the overall thermal noise in the system, making it easier to observe strong entanglement, reaching the maximum $E_{1,{\rm max}}^{(2)}\simeq 1.8$.

\subsection{Triadic entanglement}

 \begin{figure}
	\centering
	\includegraphics[width=\linewidth]{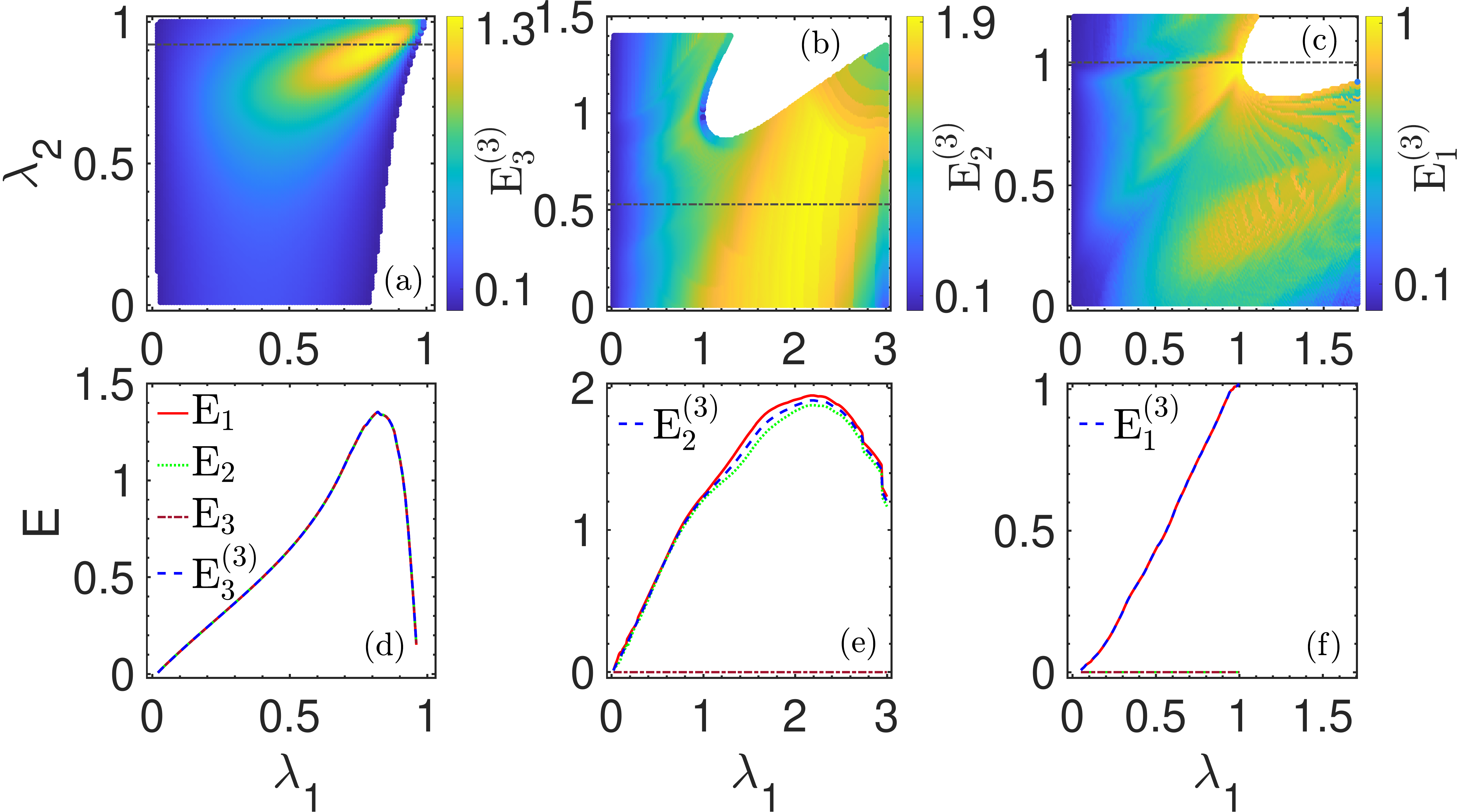}
	\caption{\label{fig3}
	Logarithmic negativity for triadic entanglement. Panels (a)--(c) show numerically optimized entanglement for (a) all bipartitions, $E_3^{(3)}$, (b) two bipartitions, $E_2^{(3)}$, and (c) one bipartition, $E_1^{(3)}$.
	Panels (d)--(f) show cuts through the corresponding 2D plots for entanglement in (d) all bipartitions for $\lambda_2 = 0.92$, (e) two bipartitions for $\lambda_2 = 0.53$, and (f) one bipartition for $\lambda_2 = 1.01$. System parameters for numerical simulations and colour coding are the same as in Fig.~\ref{fig2}.
		}
\end{figure}

We investigate triadic entanglement in the three-particle system numerically in Fig.~\ref{fig3}. We again start with entanglement in all three bipartitions (panel (a)) for which we assume equal coupling for all three Bogoliubov modes $G_1 = G_2 = G_3$. We again observe equal entanglement in all three bipartitions (panel (d) for a cut along $\lambda_2 = 0.91$). Since we are not discarding part of the system (like in the case of dyadic entanglement, where one of the three particles is traced out to calculate the logarithmic negativity), the attainable entanglement is larger with $E_{1,{\rm max}}^{(3)}\simeq 1.4$. This simultaneous entanglement in all bipartitions of the three-particle system demonstrates that generating genuine tripartite entanglement is possible with feasible experimental parameters.

Breaking the symmetry in coupling rates, $G_1\neq G_2\neq G_3$, allows us to generate strong entanglement only in two or one bipartitions. For entanglement in two bipartitions (panel (b)), the maximum can be reached, somewhat surprisingly, for strong cooling of a single Bogoliubov mode, $G_1\gg G_2\sim G_3$. This apparent paradox can be understood from the structure of the Bogoliubov mode with coefficients satisfying $\lambda_1\sim\lambda_2\gg\lambda_3$: This setting creates strong entanglement in the pair of particles $b_1$ and $b_2$, which is shared by two bipartitions. This intuition is supported by our numerical simulations which show that indeed only this pair of particles is entangled (more on the relationship between dyadic and triadic entanglement below and in Table~\ref{tab:interplay}). Both bipartitions show similar amounts of entanglement (see panel (e) showing cut along $\lambda_2 = 0.53$) with maximum $E_{2,{\rm max}}^{(3)}\simeq 1.9$.

Last but not least, entanglement in one bipartition can be prepared as well (see panel (c)). Strong cooling of one Bogoliubov mode, $G_1\gg G_2\sim G_3$, but with comparable coeffcients, $\lambda_1\sim\lambda_2\sim\lambda_3$, can best achieve this task, leading to the maximum $E_{1,{\rm max}}^{(3)}\simeq 1.0$. The surprisingly weaker entanglement (compared to entanglement in two and three bipartitions) is caused by the strong cooling of a single Bogoliubov mode which results in strong thermal noise in the remaining normal modes of the three-particle system. Since the noise properties for each particle are determined by the combination of all three Bogoliubov modes, this competition between strong entanglement in one Bogoliubov mode and thermal noise in the remaining Bogoliubov modes results in weaker entanglement than for $E_{2,3}^{(3)}$. Panel (f) again shows a cut through the maximum for $\lambda_2 = 1.01$.

\subsection{Interplay between dyadic and triadic entanglement}

\begin{table*}
	\centering
	\caption{\label{tab:interplay}Interplay of dyadic and triadic entanglement and squeezed collective quadratures. The left half shows results for maximum dyadic entanglement $E_{j,{\rm max}}^{(2)}$, including the corresponding triadic entanglement and the dyadic and triadic collective quadratures that are squeezed below the shot-noise level. The right half shows the squeezed quadratures and corresponding dyadic entanglement for the maxima of triadic entanglement $E_{j,{\rm max}}^{(3)}$. To keep the notation concise, the collective quadratures are not normalized.}
	\begin{tabular}{cp{0.5cm}cccc p{0.5cm} cccc}
		\hline
		\multirow{2}{*}{$j$} && \multirow{2}{*}{$E_{j,{\rm max}}^{(2)}$} & \multirow{2}{*}{$E_k^{(3)}$} 
			& \multicolumn{2}{c}{Squeezed quadratures} && \multirow{2}{*}{$E_{j,{\rm max}}^{(3)}$} & \multirow{2}{*}{$E_k^{(2)}$} 
			& \multicolumn{2}{c}{Squeezed quadratures} \\
			&&&& dyadic & triadic &&&& dyadic & triadic \\
		\hline
		\multirow{4}{*}{3} && \multirow{4}{*}{0.36} & \multirow{4}{*}{$E_3^{(3)}\simeq 0.98$}
			& $p_1-p_2$ & $x_1+x_2+x_3$ && \multirow{4}{*}{1.4} & \multirow{4}{*}{$E_3^{(2)}\simeq 0.027$} & $p_1-p_2$ & $x_1+x_2+x_3$ \\
			&&&& $p_2-p_3$ & $-p_1+p_2+p_3$ &&&& $p_2-p_3$ & $-p_1+p_2+p_3$ \\
			&&&& $-p_1+p_3$ & $p_1-p_2+p_3$ &&&& $-p_1+p_3$ & $p_1-p_2+p_3$ \\
			&&&&& $p_1+p_2-p_3$ &&&&& $p_1+p_2-p_3$ \\
		\hline 
		\multirow{4}{*}{2} && \multirow{4}{*}{0.42} & \multirow{4}{*}{$E_3^{(3)}\simeq 0.87$} 
			& $x_1+x_3$ & \multirow{2}{*}{$x_1+x_2+x_3$} && \multirow{4}{*}{1.9} & \multirow{4}{*}{$E_1^{(2)}\simeq 0.46$} & \multirow{2}{*}{$x_2+x_3$} & \multirow{2}{*}{$x_1+x_2+x_3$} \\
			&&&& $p_1-p_3$ \\
			&&&& $x_2+x_3$ & \multirow{2}{*}{$p_1+p_2-p_3$} &&&& \multirow{2}{*}{$p_1-p_3$} & \multirow{2}{*}{$-p_1+p_2+p_3$} \\
			&&&& $p_2-p_3$ \\
		\hline 
		\multirow{4}{*}{1} && \multirow{4}{*}{1.8} & \multirow{4}{*}{$E_2^{(3)}\simeq 1.9$}
			& \multirow{2}{*}{$x_2+x_3$} & $x_1-p_2+p_3$ && \multirow{4}{*}{1.0} & \multirow{4}{*}{---} & \multirow{2}{*}{$p_1-p_2$} & \multirow{2}{*}{$x_1+x_2+x_3$} \\
			&&&&& $x_1+p_2-p_3$ \\
			&&&& \multirow{2}{*}{$p_2-p_3$} & $x_1+x_2+x_3$ &&&& \multirow{2}{*}{$p_1-p_3$} & \multirow{2}{*}{$-p_1+p_2+p_3$} \\
			&&&&& $-x_1+x_2+x_3$ \\
		\hline
	\end{tabular}
\end{table*}

Dyadic and triadic entanglement coexist simultaneously in the system, which we summarize in Table~\ref{tab:interplay} and discuss below. When cooling all Bogoliubov modes with equal strength, $G_1 = G_2 = G_3$, entanglement in all pairs and all bipartitions is created at the same time (assuming sufficiently large cooling rate). However, the maximum of dyadic and triadic entanglement is achieved for different Bogoliubov coefficients as can be seen by comparing Figs.~\ref{fig2} and \ref{fig3}. The coexistence of dyadic and triadic entanglement in these genuinely tripartite entangled states also affects the collective quadratures that are squeezed below the vacuum level: Only differences in momenta, $(p_i-p_j)/\sqrt{2}$ with $i,j=1,2,3$ and $i\neq j$, are squeezed for dyadic entanglement while sums of positions, $(x_i+x_j)/\sqrt{2}$, remain above the vacuum noise level. In collective quadratures of all three particles, the sum of all positions, $(x_1+x_2+x_3)/\sqrt{3}$, and collective momenta of the form $(-p_i+p_j+p_k)/\sqrt{3}$ with cycling permutations of $i\neq j\neq k$ exhibit squeezing.

When the Bogoliubov modes are cooled with unequal strengths to create entanglement in two pairs of particles, all three bipartitions remain inseparable for triadic entanglement. At the maximum two-pair dyadic entanglement $E_{2,{\rm max}}^{(2)}$, the two pairs exhibit squeezing in the sum of positions and difference of momenta of the two particles in the pair. On the other hand, despite entanglement in all three bipartitions for the triadic entanglement, only the three-particle quadratures $(x_1+x_2+x_3)/\sqrt{3}$ and $(p_1+p_2-p_3)/\sqrt{3}$ show squeezing. Uncovering the full set of squeezed collective quadratures would require optimizing the general Bogoliubov modes $\sum_i(a_ix_i+b_ip_i)$ over the coefficients $a_i,b_i$ to find the minimum variance. While such an analysis is possible, its details depend on the details of the specific experimental configuration, so we omit it here.

Generation of entanglement in two triadic bipartitions leads to dyadic entanglement in the one pair of particles that is shared by both particles---for example, when entangling the bipartitions $b_1| b_2b_3$ and $b_3| b_1b_2$, the pair of particles $b_1$ and $b_3$ becomes entangled as well. This pair of particles also shows squeezing in the sum of positions and difference of momenta. Nevertheless, squeezing in three-particle quadratures depends on the chosen system parameters. For the maximum of one-pair dyadic entanglement (Fig.~\ref{fig2}(c) where the bipartitions $b_2|b_3b_1$ and $b_3|b_1b_2$ are entangled) the following collective quadratures are squeezed: $(x_1-p_2+p_3)\sqrt{3}$, $(x_1+p_2-p_3)/\sqrt{3}$, $(x_1+x_2+x_3)/\sqrt{3}$, $(-x_1+x_2+x_3)/\sqrt{3}$. On the other hand, when maximizing entanglement in triadic bipartitions (Fig.~\ref{fig3}(b) where the bipartitions $b_1|b_2b_3$ and $b_3|b_1b_2$ are entangled), only the quadratures $(x_1+x_2+x_3)/\sqrt{3}$ and $(-p_1+p_2+p_3)/\sqrt{3}$ show squeezing.

Finally, when a single triadic bipartition is entangled, no dyadic entanglement can be observed. In the full three-particle state, the collective quadratures that are squeezed below the vacuum level are then the sum of positions and difference of momenta corresponding to the bipartition being entangled---for the case $G_1\gg G_2\sim G_3$ considered above (which entangles the particle $b_1$ with the two-particle subsystem consisting of $b_2$ and $b_3$), the squeezed quadratures are $(x_1+x_2+x_3)/\sqrt{3}$ and $(-p_1+p_2+p_3)/\sqrt{3}$. Even though the system does not exhibit dyadic entanglement, the differences of momenta of two particles, $(p_i-p_j)/\sqrt{2}$, become squeezed.

\section{Discussion}

\subsection{Experimental considerations}

The parameters for our numerical simulation are consistent with recent experiments demonstrating control of levitated nanoparticles via coherent scattering \cite{delic2019, Windey2019, delic2020}. To enable efficient coupling of each particle to three cavity modes that is comparable with existing experiments (aimed at coupling the motion to a single cavity mode), we assume that the sum of the three tweezers for each particle gives rise to a total harmonic potential with frequency of about \SI{300}{\kilo\hertz}. The relative strength of the three tweezers is then chosen to give the required coupling to each of the three cavity modes.

The biggest difference between our parameters and recent experiments is the decoherence rate of the particle motion. The assumed mechanical quality factor $Q = \omega_m/\gamma = 5\times 10^9$ and thermal occupation $\bar{n} = 2\times 10^7$ (corresponding to a \SI{300}{\kilo\hertz} mode at room temperature) gives rise to thermalisation rate $\gamma\bar{n} = 2\pi\times\SI{1.2}{\kilo\hertz}$. The total heating rate in Ref.~\cite{delic2020} consists of gas damping at rate \SI{16}{\kilo\hertz} (which can be reduced by improving the vacuum below \SI{e-6}{\milli\bar}) and photon recoil of \SI{6}{\kilo\hertz}. This recoil heating rate can be improved by using longer optical wavelength for trapping, heavier particles, and higher mechanical frequencies~\cite{Delic2020a}; fully understanding and improving nanoparticle thermalization in optical traps is one of the main topics of research in the field~\cite{Jain2016}.

Scaling to larger numbers of particles poses several critical challenges. On the theoretical side, systematic analysis of entanglement in four or more particles requires tools beyond the positive partial transpose, which is necessary and sufficient only for $1\times M$ bipartite Gaussian states; the PPT criterion therefore cannot fully characterise entanglement in four particles since it is not a necessary condition for $2\times 2$ Gaussian systems. In addition, such an approach is viable only for small particle arrays as the number of possible bipartitions grows exponentially with the array size. Efficient classification of entanglement in larger arrays could benefit from advanced numerical techniques based on machine learning~\cite{Carleo2019} which have, however, been developed only for tomography of single-particle states and dynamics so far~\cite{Conangla2019,Weiss2019}.

On the experimental side, dissipative generation of a general $N$-particle state requires cooling of $N$ collective Bogoliubov modes. This strategy requires splitting the total trapping potential for each particle into $N$ trapping beams separated in frequency by integer multiples of the free spectral range of the optical cavity. Combining all these tweezers for each particle efficiently represents a significant technological challenge. Experiments with large nanoparticle arrays therefore require novel strategies for efficient preparation of complex multiparticle entangled states with minimal resources. While our analysis offers only a glimpse into these issues, we believe such approaches to be feasible with modest theoretical efforts (see Methods).

Finally, to avoid the need for a large number of optical cavity modes to remove thermal noise from long nanoparticle arrays, coherent scattering can be complemented with measurement feedback~\cite{Zhang2017}. Measurement-based feedback has recently been used to cool nanoparticle motion to the quantum ground state~\cite{Magrini2021,Tebbenjohanns2021} and this approach can be scaled to multiple particles~\cite{Vijayan2022X}. When combined with direct coupling between nanoparticles (mediated by the Coulomb force), it can be used to create steady-state entanglement as well~\cite{Rudolph2022X}. Feedback techniques generally operate without a cavity which would only reduce the measurement bandwidth, leading to a reduced rate at which information about the state of the particle is acquired. Feedback can then be used to reduce the thermal noise in the whole nanoparticle array while only a few Bogoliubov modes are cooled via coherent scattering to create the desired entangled state; using coherent scattering guarantees tuneability of the generated entanglement without undesired crosstalk in direct inter-particle interactions needed to generate entanglement via feedback~\cite{Rudolph2022X}.

\subsection{Conclusions}

In summary, we proposed and analysed a deterministic scheme to generate and control entanglement in levitated nanoparticle arrays. Applying multiple tweezers per particle and scattering photons coherently into separate cavity modes allows us to cool suitably engineered collective Bogoliubov modes, leading to strong Gaussian entanglement between nanoparticles. The general structure of Bogoliubov modes we considered allows great tuneability of the resulting bipartite entanglement, opening the way to a range of applications of entangled levitated nanoparticles in fundamental physics and quantum technologies.

To fully characterise the generated entanglement and provide an essential stepping stone for future theoretical and experimental efforts, we focused on the case of three particles, for which the positive partial transpose provides a necessary and sufficient condition for separability. We showed how limiting ourselves to permutation symmetric Bogoliubov modes allows tuning between entanglement in one, two, or three bipartitions simply by changing the coefficients in the Bogoliubov modes and their coupling rates, while keeping the resulting entanglement structure easy to understand. 
In the future, further optimization of the generated entanglement can be achieved by considering more general (i.e., not permutationally invariant) Bogoliubov modes. Given the large parameter space available already for three particles, machine learning approaches would be particularly beneficial, allowing also scaling of our strategy to larger nanoparticle arrays. Such arrays could then serve as an important testing ground for investigating the relationship between the structure of Bogoliubov modes and the resulting multiparticle entanglement and other optomechanical many-body experiments~\cite{Xuereb2014}.

Introducing direct coupling between levitated nanoparticles (possible via Coulomb interactions between charged particles~\cite{Frimmer2017} or by optically mediated dipole forces~\cite{rieser2022,DeBernardis2022}) would further enhance the capabilities of levitated nanoparticle arrays and allow us to use them as a platform for simulating quantum many-body models~\cite{Bernien2017,Georgescu2014}. This step would extend the applications of levitated nanoparticles from direct tests of fundamental physics~\cite{Romero-Isart2011,Pontin2020} and searching for new physics~\cite{Moore2014,Rider2016,Monteiro2020} to other quantum technology applications.

With the recent progress in quantum control of optically trapped nanoparticles and growing theoretical and experimental interest in multiparticle systems, our work presents an important step to increasing the number of particles, enabling the development of quantum sensing protocols and many-body experiments in optically levitated nanoparticles. As a technique borrowed from atomic physics, coherent scattering shows great promise as a tool for controlling arrays of levitated nanoparticles; the protocols developed within the context of levitated optomechanics can, in turn, inspire future theoretical and experimental efforts in the many-body dynamics of atomic arrays in optical lattices. As a tool for controlling both levitated particles and single atoms, coherent scattering can thus open the door to experiments involving both types of systems where the large mass imbalance can be used to observe new physical effects~\cite{Jockel2015}.

\section{Methods}

\subsection{Lyapunov equation}\label{app:Lyapunov}

To find the steady state of the optomechanical system, we start from the Langevin equations for the quadrature operators written in the matrix form,
\begin{equation}
	\dot{r} = Ar+\xi.
\end{equation}
Here $r = (X_1,Y_1,\ldots X_N,Y_N,x_1,p_1,\ldots, x_N,p_N)^T$ is the vector of quadrature operators, $\xi = (\sqrt{\kappa_1}X_{1,{\rm in}},\sqrt{\kappa_1}Y_{1,{\rm in}}, \ldots,\sqrt{\gamma_N}x_{N,{\rm in}},\sqrt{\gamma_N}p_{N,{\rm in}})^T$ are the corresponding input noises, and
\begin{equation}
	A = \begin{pmatrix}
		A_c & A_{cm} \\ A_{mc} & A_m
	\end{pmatrix}
\end{equation}
is the drift matrix written in terms of the blocks describing the dynamics of the cavity modes, mechanical modes, and their interactions,
\begin{subequations}
\begin{align}
	A_c &= -\frac{1}{2}{\rm diag}(\kappa_1,\kappa_1,\ldots,\kappa_N,\kappa_N), \\
	A_m &= -\frac{1}{2}{\rm diag}(\gamma_1,\gamma_1,\ldots,\gamma_N,\gamma_N), \\
	A_{cm} &= \begin{pmatrix}
		A_{11} & \ldots & A_{N1} \\
		\vdots &\ddots & \\
		A_{1N} & \ldots & A_{NN}
	\end{pmatrix} \\
	A_{mc} &= \begin{pmatrix}
		A_{11} & \ldots & A_{1N} \\
		\vdots &\ddots & \\
		A_{N1} & \ldots & A_{NN}
	\end{pmatrix},
\end{align}
\end{subequations}
where we introduced the $2\times2$ blocks $A_{jk}$ of the form
\begin{equation}
	A_{jk} = \begin{pmatrix}
		0 & g_{-,jk}-g_{+,jk} \\ -g_{-,jk}-g_{+,jk} & 0
	\end{pmatrix}.
\end{equation}
The steady state of the system can now be expressed in terms of the covariance matrix which obeys the Lyapunov equation
\begin{equation}
	{AV}+{VA}^{T}+ {N}=0,
\end{equation}
where ${V}$ with elements $V_{jk}=\langle r_{j}r_{k} +r_{k}r_{j} \rangle-2\langle r_{j}\rangle \langle r_{k}\rangle$ is the covariance matrix and $N = \langle\xi(t)\xi^T(t)\rangle = {\rm diag}[\kappa_1,\kappa_1,\ldots,\kappa_N,\kappa_N,\gamma_1(2\bar{n}+1),\gamma_1(2\bar{n}+1),\ldots,\gamma_N(2\bar{n}+1),\gamma_N(2\bar{n}+1)]$ is the diffusion matrix.

For $N=3$ and the Bogoliubov modes given by Eqs.~\eqref{M_b2}, the coupling rates $g_{\pm,jk}$ can be determined from the coefficients $\lambda_j$ and coupling of the Bogoliubov modes $G_k$. Writing the Hamiltonian~\eqref{M_b1} in terms of the mechanical modes $b_j$, we directly obtain
\begin{align}
\begin{split}
	g_{+,11} = \lambda_1 G_1,\quad g_{-,21} = \lambda_2 G_1,\quad g_{-,31} = \lambda_3 G_1,\\
	g_{-,12} = \lambda_3 G_2,\quad g_{+,22} = \lambda_1 G_2,\quad g_{-,32} = \lambda_2 G_2,\\
	g_{-,13} = \lambda_2 G_1,\quad g_{-,23} = \lambda_3 G_2,\quad g_{+,33} = \lambda_1 G_3;
\end{split}
\end{align}
all remaining coefficients are zero.

\subsection{Cooling non-orthogonal Bogoliubov modes}

When cooling non-orthogonal Bogoliubov modes, it is generally not possible to cool them all to their quantum ground states. This can be illustrated on the case of two Bogoliubov modes parametrized as
\begin{subequations}
\begin{align}
	\beta_1 &= u_1b_1^\dagger +u_2b_2,\\
	\beta_2 &= v_1b_1 + v_2b_2^\dagger,
\end{align}
\end{subequations}
where we assume $u_j,v_j\in\mathbb{R}$ for simplicity. If the first Bogoliubov mode is in its quantum ground state, $\langle\beta_1^\dagger\beta_1\rangle = 0$, the inter-particle correlation $\langle b_1b_2+b_1^\dagger b_2^\dagger\rangle$ is determined by the occupations of the particle modes $n_j = \langle b_j^\dagger b_j\rangle$ via the relation (assuming $u_{1,2}\neq 0$)
\begin{equation}
	\langle b_1b_2+b_1^\dagger b_2^\dagger\rangle = -\frac{u_1^2(n_1+1)+u_2^2n_2}{u_1u_2}.
\end{equation}
With this result, the occupation of the second Bogoliubov mode is given by
\begin{equation}
	\langle\beta_2^\dagger\beta_2\rangle = v_1^2\left(1-\frac{u_1v_2}{u_2v_1}\right)n_1 + v_2^2\left(1-\frac{u_2v_1}{u_1v_2}\right)n_2 + v_2^2\left(1-\frac{u_1v_1}{u_2v_2}\right).
\end{equation}
Since $n_j\geq 0$, zero occupation of the Bogoliubov mode $\beta_2$ is possible only when $u_1v_1 = u_2v_2$ which implies orthogonal modes; we then have the commutator
\begin{equation}
	[\beta_1,\beta_2] = -u_1v_1+u_2v_2 = 0.
\end{equation}
In addition, even for orthogonal Bogoliubov modes, we can have $\langle\beta_2^\dagger\beta_2\rangle = 0$ only if $u_1 = u_2 = v_1 = v_2$ which corresponds to a maximally entangled (and infinitely squeezed) state.

\subsection{Genuine tripartite entanglement from two-particle Bogoliubov modes}

To estimate the feasibility of creating strong multipartite entanglement between $N$ particles from few-particle Bogoliubov modes, we analyze generation of genuine tripartite entanglement from two-particle Bogoliubov modes. We set $\lambda_3=0$ in Eq.~\eqref{M_b2}, obtaining the Bogoliubov modes
\begin{equation}
	\beta_1 = \lambda_1 b_1^\dagger + \lambda_2 b_2,\ \beta_2 = \lambda_1 b_2^\dagger + \lambda_2 b_3,\ \beta_3 = \lambda_1 b_3^\dagger + \lambda_2 b_1
\end{equation}
with $-\lambda_1^2+\lambda_2^2 = 1$.
With this set of modes, we aim to maximize the genuine tripartite entanglement $E_3^{(3)}$ by optimizing the coupling rates $G_k$.

\begin{figure}
	\centering
	\includegraphics[width=\linewidth]{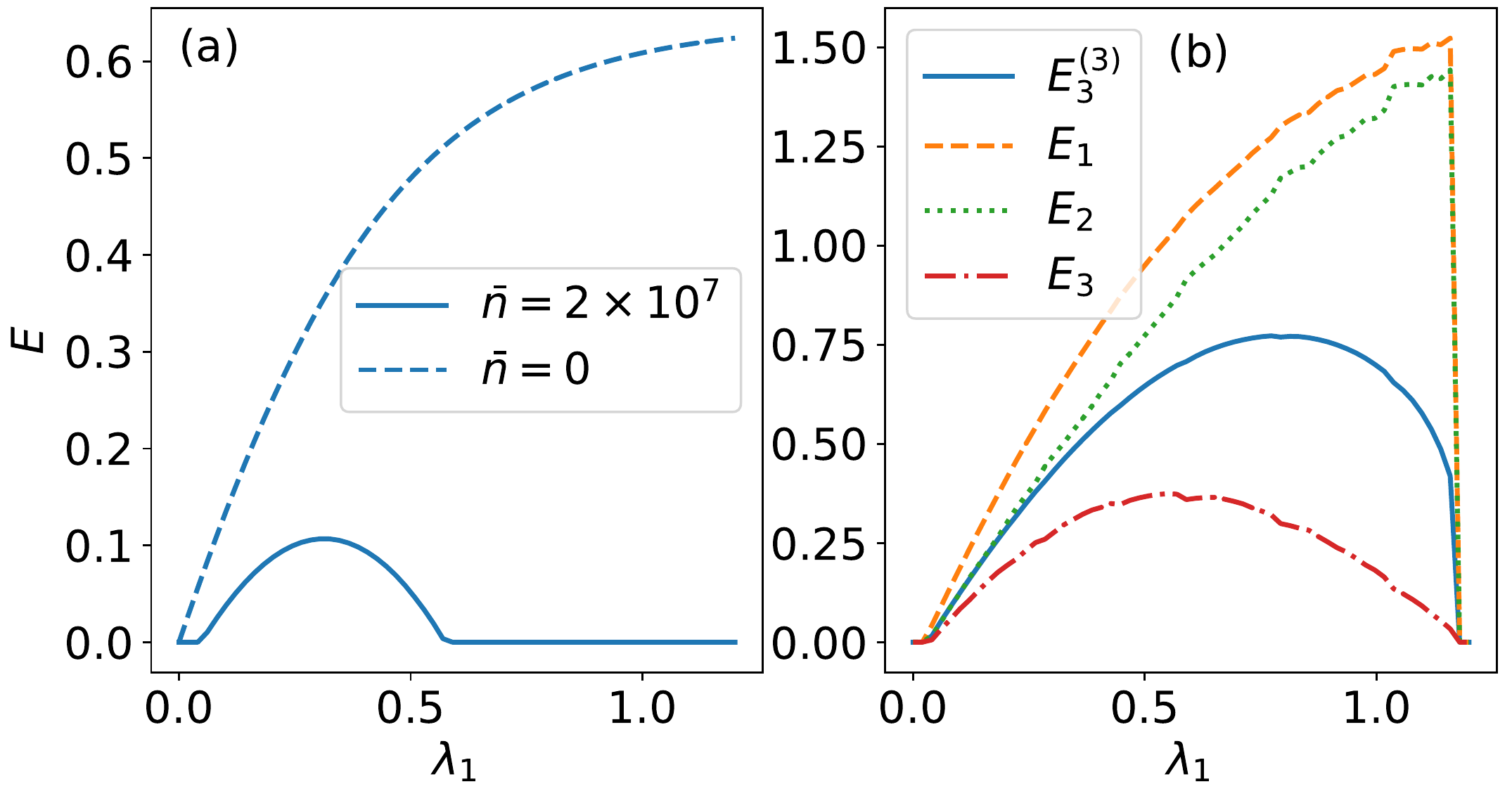}
	\caption{\label{fig:Bog2}
		Genuine tripartite entanglement with two particles per Bogoliubov mode. (a) Entanglement $E_3^{(3)}$ optimized over symmetric coupling $G_1=G_2=G_3$ in the presence of thermal noise ($\bar{n}=2\times 10^7$, solid) and in its absence ($\bar{n}=0$, dashed). (b) Entanglement $E_3^{(3)}$ (solid blue line) optimized over general coupling rates $G_1\neq G_2\neq G_3$. The asymmetry in coupling the three Bogoliubov modes to the cavity fields results in unequal entanglement in the three bipartitions as shown by plotting the corresponding logarithmic negativities $E_{1,2,3}$. System parameters are the same as in Fig.~\ref{fig2}.}
\end{figure}

The results of this optimization are shown in Fig.~\ref{fig:Bog2}. In panel (a), we study entanglement generation with equal coupling rates $G_1=G_2=G_3$, resulting in equal entanglement in all bipartitions. For realistic experimental parameters, the attainable entanglement remains weak, reaching the maximum $E_{3,{\rm max}}^{(3)}\simeq 0.11$ (solid line); however, reducing thermal decoherence of the nanoparticle motion has the potential to significantly increase the amount of entanglement (see dashed line for zero-temperature mechanical bath, $\bar{n}=0$).
Further improvements are possible with asymmetric coupling of the three Bogoliubov modes, $G_1\neq G_2\neq G_3$, as shown in Fig.~\ref{fig:Bog2}(b). The entanglement is now distributed unequally among the three bipartitions but even the most weakly entangled bipartition (dot-dashed red line) can reach entanglement of $E_{3,{\rm max}}\simeq 0.37$ which surpasses the entanglement attainable for symmetric coupling.

%

\section{Acknowledgements}
	We would like to thank Uro\v{s} Deli\'{c} for useful discussions. We gratefully acknowledge financial support by the project 20-16577S of the Czech Science Foundation and projects No. CZ.02.1.01/0.0/0.0/16\textunderscore{}026/0008460 and 8J21AT007 of the Czech Ministry of Education, Youth and Sports (MEYS \v{C}R).

\section{Author contributions}

O.\v{C}. and R.F. developed the idea. A.K.C. performed analytical and numerical calculations and interpreted the results with all authors. A.K.C. and O.\v{C}. wrote the manuscript with input from R.F who supervised the project.

\section{Competing interests}

The authors declare no competing interests.

\end{document}